\documentclass{wscpaperproc}
\usepackage{latexsym}
\usepackage{graphicx}
\usepackage{mathptmx}
\usepackage[T1]{fontenc}

\usepackage{amsmath}
\usepackage{amsfonts}
\usepackage{amssymb}
\usepackage{amsbsy}
\usepackage{amsthm}
\usepackage{booktabs}
\usepackage[shortlabels]{enumitem}
\usepackage{multicol}
\usepackage{multirow}

\usepackage[pdftex,colorlinks=true,urlcolor=blue,citecolor=black,anchorcolor=black,linkcolor=black]{hyperref}
\usepackage{cleveref}

\newtheoremstyle{wsc}
{3pt}
{3pt}
{}
{}
{\bf}
{}
{.5em}
{}

\theoremstyle{wsc}

\begin{document}

%
%

\pagestyle{fancyplain}

\thispagestyle{plain}
\firstPageHead{}

\chead{\fancyplain{}{\itshape Chen, Wu, McCartney, Sadeh and Fang}}

\rhead{}
\cfoot{}
\renewcommand{\headrulewidth}{0pt} 

\makeatletter
\let\@internalcite\cite
\def\cite{\def\@citeseppen{-1000}%
    \def\@cite##1##2{(##1\if@tempswa , ##2\fi)}%
    \def\citeauthoryear##1##2##3{##1 ##3}\@internalcite}
\def\citeNP{\def\@citeseppen{-1000}%
    \def\@cite##1##2{##1\if@tempswa , ##2\fi}%
    \def\citeauthoryear##1##2##3{##1 ##3}\@internalcite}
\def\citeN{\def\@citeseppen{-1000}%
    \def\@cite##1##2{##1\if@tempswa, ##2)\else{}\fi}%
    \def\citeauthoryear##1##2##3{##1 (##3)}\@citedata}
\def\citeA{\def\@citeseppen{-1000}%
    \def\@cite##1##2{(##1\if@tempswa , ##2\fi)}%
    \def\citeauthoryear##1##2##3{##1}\@internalcite}
\def\citeANP{\def\@citeseppen{-1000}%
    \def\@cite##1##2{##1\if@tempswa , ##2\fi}%
    \def\citeauthoryear##1##2##3{##1}\@internalcite}
\def\shortcite{\def\@citeseppen{-1000}%
    \def\@cite##1##2{(##1\if@tempswa , ##2\fi)}%
    \def\citeauthoryear##1##2##3{##2 ##3}\@internalcite}
\def\shortciteNP{\def\@citeseppen{-1000}%
    \def\@cite##1##2{##1\if@tempswa , ##2\fi}%
    \def\citeauthoryear##1##2##3{##2 ##3}\@internalcite}
\def\shortciteN{\def\@citeseppen{-1000}%
    \def\@cite##1##2{##1\if@tempswa, ##2\else{}\fi}%
    \def\citeauthoryear##1##2##3{##2 (##3)}\@citedata}
\def\shortciteA{\def\@citeseppen{-1000}%
    \def\@cite##1##2{(##1\if@tempswa , ##2\fi)}%
    \def\citeauthoryear##1##2##3{##2}\@internalcite}
\def\shortciteANP{\def\@citeseppen{-1000}%
    \def\@cite##1##2{##1\if@tempswa , ##2\fi}%
    \def\citeauthoryear##1##2##3{##2}\@internalcite}
\def\citeyear{\def\@citeseppen{-1000}%
    \def\@cite##1##2{(##1\if@tempswa , ##2\fi)}%
    \def\citeauthoryear##1##2##3{##3}\@citedata}
\def\citeyearNP{\def\@citeseppen{-1000}%
    \def\@cite##1##2{##1\if@tempswa , ##2\fi}%
    \def\citeauthoryear##1##2##3{##3}\@citedata}
%
%
%
\def\@citedata{%
    \@ifnextchar [{\@tempswatrue\@citedatax}%
                  {\@tempswafalse\@citedatax[]}%
}

\def\@citedatax[#1]#2{%
\if@filesw\immediate\write\@auxout{\string\citation{#2}}\fi%
  \def\@citea{}\@cite{\@for\@citeb:=#2\do%
    {\@citea\def\@citea{, }\@ifundefined
       {b@\@citeb}{{\bf ?}%
       \@warning{Citation `\@citeb' on page \thepage \space undefined}}%
{\csname b@\@citeb\endcsname}}}{#1}}%

%
\def\@citex[#1]#2{%
\if@filesw\immediate\write\@auxout{\string\citation{#2}}\fi%
  \def\@citea{}\@cite{\@for\@citeb:=#2\do%
    {\@citea\def\@citea{; }\@ifundefined
       {b@\@citeb}{{\bf ?}%
       \@warning{Citation `\@citeb' on page \thepage \space undefined}}%
{\csname b@\@citeb\endcsname}}}{#1}}%

%
\def\@biblabel#1{}
\makeatother



\newdimen\bibindent
\bibindent=0.0em
\def\thebibliography#1{\section*{\refname}\list
   {}{\settowidth\labelwidth{[#1]}
   \leftmargin\parindent
   \itemindent -\parindent
   \listparindent \itemindent
   \itemsep 0pt
   \parsep 0pt}
   \def\newblock{}
   \sloppy
   \sfcode`\.=1000\relax}


\setlength{\baselineskip}{12.7pt}

\title{OUT OF THE PAST: AN AI-ENABLED PIPELINE FOR TRAFFIC SIMULATION FROM NOISY, MULTIMODAL DETECTOR DATA AND STAKEHOLDER FEEDBACK}

\author{\begin{center}
Rex Chen\textsuperscript{1}, 
Karen Wu\textsuperscript{1}, 
John McCartney\textsuperscript{2}, 
Norman Sadeh\textsuperscript{1}, and 
Fei Fang\textsuperscript{1}\\
[11pt]
\textsuperscript{1}School of Computer Science, Carnegie Mellon University, Pittsburgh, PA, UNITED STATES\\
\textsuperscript{2}Path Master Inc., Twinsburg, OH, UNITED STATES
\end{center}
}

\maketitle

\vspace{-12pt}

\section*{Abstract}
How can a traffic simulation be designed to faithfully reflect real-world traffic conditions? One crucial step is modeling the volume of traffic demand. But past demand modeling approaches have relied on unrealistic or suboptimal heuristics, and they have failed to adequately account for the effects of noisy and multimodal data on simulation outcomes. In this work, we integrate advances in AI to construct a three-step, end-to-end pipeline for systematically modeling traffic demand from detector data: computer vision for vehicle counting from noisy camera footage, combinatorial optimization for vehicle route generation from multimodal data, and large language models for iterative simulation refinement from natural language feedback. Using a road network from Strongsville, Ohio as a testbed, we show that our pipeline accurately captures the city's traffic patterns in a granular simulation. Beyond Strongsville, incorporating noise and multimodality makes our framework generalizable to municipalities with different levels of data and infrastructure availability.

\section{Introduction}
\label{sec:intro}
Traffic simulation is an important tool in transportation research: both for performing traffic analysis on experimental substitutes of real-world transportation systems \shortcite{Barcelo2010}, and for training and evaluating intelligent transportation systems, e.g., those based on reinforcement learning \shortcite{Mei2024}. If the results of traffic simulations are to be deployed in real-world transportation systems, they must be sufficiently realistic to foster trust from stakeholders. Although there is a significant body of work on constructing efficient \emph{simulators} for executing simulations \shortcite{Zhang2019,Chen2023}, an understudied problem is the construction of realistic \emph{simulations} grounded in data from physical traffic systems.

Existing approaches to creating road network-scale traffic simulations have a number of limitations that hamper their realism and thus their practical applicability. We focus on limitations surrounding the central task of \emph{demand modeling}, or the modeling of traffic volumes within the simulation. Demand modeling methods that construct origin-destination matrices from activity data are unrealistic and fail to make use of traffic detector data. Meanwhile, detector data-driven approaches to demand modeling have relied on suboptimal heuristics. All of these approaches also consider the source data to be the ground truth; they do not perform any calibration to account for sources of noise or multimodality in the data.

\begin{figure}[ht]
    \centering
    \includegraphics[trim=1.5cm 4cm 0.5cm 0cm,clip,width=0.95\linewidth]{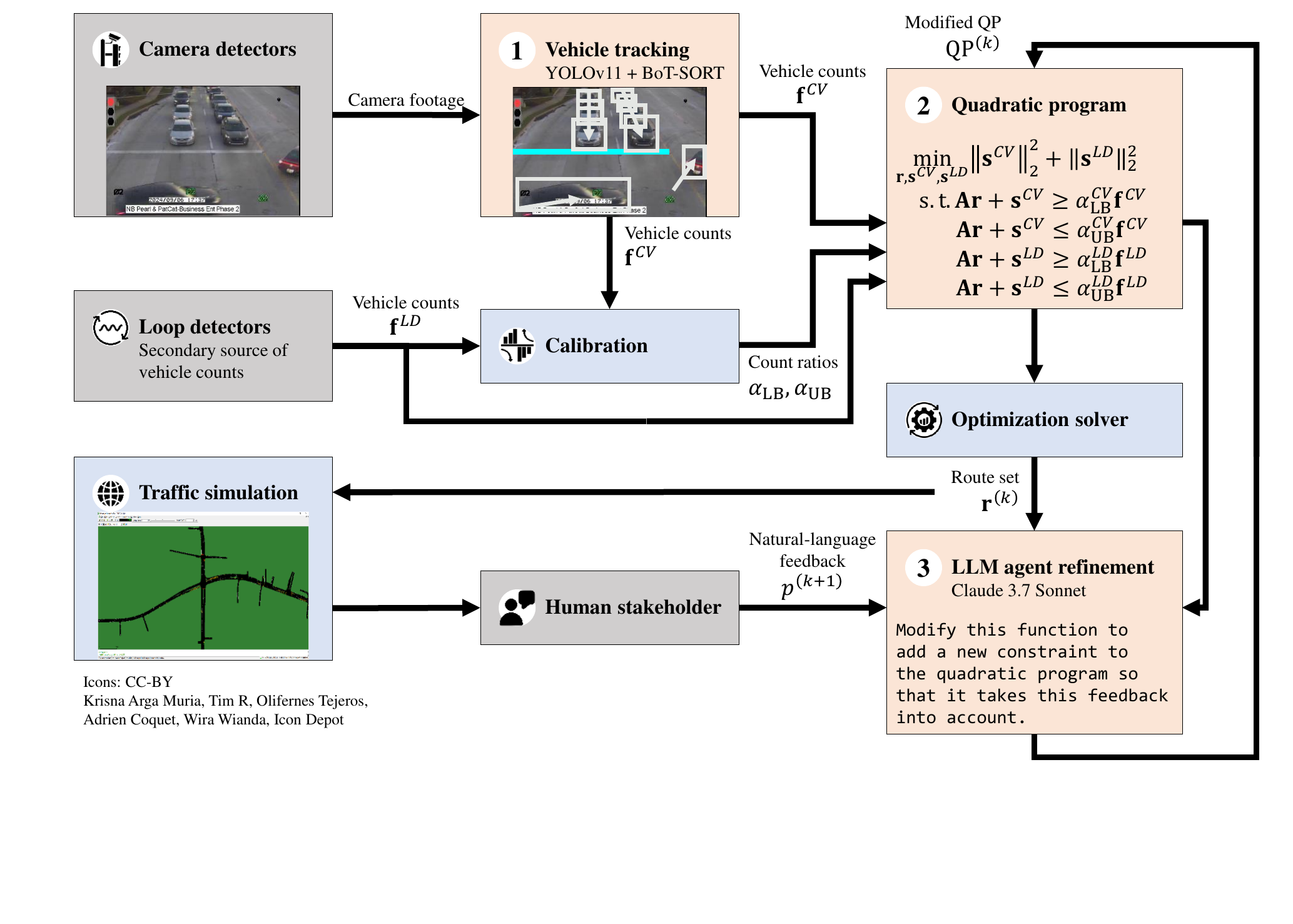}
    \caption{Our pipeline for generating a traffic simulation from multimodal detector data.}
    \label{fig:pipeline}
\end{figure}

In this work, we contribute a detailed, systematic pipeline for modeling demand in a traffic simulation from noisy, multimodal detector data (\Cref{fig:pipeline}). Starting from raw detector data, our pipeline consists of three steps: (1) We apply a vehicle tracking-based computer vision method directly to camera footage, to obtain more accurate vehicle counts than the camera detectors themselves. (2) We solve a quadratic optimization program to populate our simulation with vehicle routes. In doing so, we account for multimodality by imposing multiple sets of optimization constraints based on different sources of vehicle counts. (3) We incorporate feedback from stakeholders to refine the simulation, using a large language model (LLM) agent that encodes natural language feedback into optimization constraints.

\newpage
As a proof of concept, we apply our pipeline to simulate a high-traffic road network from the city of Strongsville, Ohio. Beginning with 24 hours of recorded camera footage and detector data from 36 intersections, we created a fully realized traffic simulation. We show that: (1) Our vehicle tracking-based computer vision method rectifies undercounting in camera detector data. (2) Our optimization method is able to generate a set of vehicle routes that is consistent with counts from both computer vision and loop detector counts, while still accounting for error in these counts. (3) Our LLM agent is able to synthesize code representing sensible, quantified constraints based on qualitative stakeholder feedback. Our pipeline code and LLM prompts are available at \url{https://github.com/lythronaxargestes/strongsville-trafficsim-public}.

\section{Related Work}
\label{sec:related}

\subsection{Demand Modeling for Traffic Simulation}
\label{sec:related:demand}
\emph{Demand modeling}, or the modeling of trips taken by individual vehicles from one point to another in a road network, is a central but difficult aspect of constructing data-driven traffic simulations \shortcite{Barcelo2010}. One popular approach is \emph{activity modeling}, where trips are extrapolated from censuses of the daily activities taken by a sample of households in the study region \shortcite{Bochenina2023,Codeca2015,Codeca2017,He2024,Leon2023,Uppoor2012}. While many municipalities collect this data for transportation planning, the locations of these activities are usually coarsely discretized. Furthermore, they represent a small, not necessarily representative sample of the population. This means that activity modeling-based simulations are prone to significant error \shortcite{Kwak2012}; we do not consider activity modeling in this chapter. 

An alternative approach to demand modeling directly uses data from traffic detectors. The types of detectors used for simulations include induction loops \shortcite{Bieker2014} and video cameras \shortcite{Wei2019,Xu2021,Zheng2019}. Detectors provide granular vehicle counts local to individual intersections, but converting them to fully-realized routes through a road network is nontrivial. \shortciteN{Wei2019,Xu2021}, and \shortciteN{Zheng2019} all provided no details on how they generated vehicle routes. The route generation procedures that have been specified in prior work have relied on suboptimal heuristics. \shortciteN{Lobo2020} and \shortciteN{Rapelli2022} used detector data to adjust activity models. \shortciteN{Bieker2014} generated routes probabilistically by using turn ratios to define distributions over movements at intersections. This approach only leads to correct simulation outcomes in expectation. Finally, \shortciteN{Qiu2024}'s approach, which uses scripts included with the traffic simulator SUMO \shortcite{AlvarezLopez2018}, is most similar to ours: they applied a two-step process of first sampling routes randomly, and then solving a linear program (LP) to approximate how many times each route should be used to match the detector counts as closely as possible. Unlike them, we solve the problem exactly as a quadratic integer program (QIP) without intermediate approximations.

One further limitation of previous detector-based approaches is that they rely on a single source of detector data, which is assumed to be generally error-free. Among the works cited previously, only \shortciteN{Bieker2014} reported detectors that failed to report vehicle counts and were removed from their dataset. When these works validate their simulations, they treat the detector data as the ground truth to compare their simulations against \shortcite{Bieker2014,Codeca2015,Leon2023,Lobo2020,Uppoor2012}. By contrast, our approach (\Cref{sec:pipeline:qp}) integrates multiple methods of processing detector data, and we adopt a semi-automated approach to validation that combines detector data with manual verification. This approach allowed us to obtain higher realism in the vein of traffic simulations that rely exclusively on manual counting \shortcite{Mueller2021}, but at a much larger scale.

\subsection{Computer Vision for Traffic Footage}
\label{sec:related:cv}
\emph{Vehicle counting} can be decomposed into two distinct but related problems: \emph{vehicle detection}, the identification of vehicles in footage; and \emph{vehicle tracking}, the identification of these vehicles' trajectories across frames \shortcite{Tituana2022}. These methods can be divided into two distinct waves of research. First, in the 1990s, advances in image processing led to vehicle detection methods based on extracting heuristically designed features \shortcite{Coifman1998}. Second, in the 2010s, the advent of \emph{convolutional neural networks} (CNNs) led to various deep object detection algorithms capable of automatically extracting relevant features for vehicle detection \shortcite{Wang2019}. While these lines of work have more recently overlapped methodologically, they have not been comparatively evaluated to our knowledge. We provide an in-situ evaluation of AutoScope, a widely-deployed image processing method, against CNN-based counting.

\subsection{Large Language Models for Transportation Research}
\label{sec:related:llm}
\emph{Large language models} (LLMs) are useful for aligning AI systems with human intuition. As such, they have been increasingly applied to the domain of transportation. One line of work has generated simulation scenarios \shortcite{Chang2024,Li2024,Sun2024} and reward functions for vehicular agents \shortcite{Han2024,Ziegler2019} based on natural language prompts. Another line of work has used LLMs to align simulations with reality based on general knowledge; \shortciteN{Da2024} used an LLM to infer how actions taken in a simulated environment would affect a real-world environment differently. 

When responding to prompts, LLMs can use external tools. For instance, \shortciteN{Li2024}'s agent generates and executes command-line calls to the SUMO simulator; \shortciteN{Wang2024}'s agent chooses between perception and decision tools to perform reinforcement learning for traffic signal control. We leverage a strength of LLMs that has not been explored for transportation research to my knowledge: the synthesis of syntactically and semantically correct programs \shortcite{Austin2021}. We use an LLM within an iterative framework in which the \emph{only} external input required is qualitative natural language feedback.

\section{Traffic Detection in Practice}
\label{sec:detectors}
Modern traffic systems make use of different types of detectors, which have varying strengths and limitations. To build a realistic traffic simulation from detector data, it is important to understand the primary use case of the data and the circumstances under which it may diverge from the ground truth. We consider two types of detectors that are commonly used in modern traffic systems.

\paragraph{Camera Detectors} 
A camera detector is typically mounted in a fixed position above the roadway, and detection zones are placed on the camera's field of view. In the United States, the AutoScope vehicle detection algorithm \shortcite{Michalopoulos1991} is used for many cameras. It extracts features to label each detection zone as being in one of three discrete states: ``background'', ``uncertain'', and ``vehicle''. While this algorithm is able to generate vehicle counts, the counts do not reflect the actual vehicle volume --- they are the number of times each zone was in the ``vehicle'' state. In high-volume traffic, consecutive vehicles may continuously actuate a detection zone (\Cref{fig:cv}a), leading to undercounting. This is more significant of an issue for traffic simulation than for the detectors' primary use in traffic signal control.

Inclement environmental conditions also contribute to inaccuracy in camera detector counts. Darkness (due to nighttime or fog) and precipitation (such as rain or snow, which cause glare) obfuscate the visual signal of vehicles, thus making it more difficult for vehicle detection algorithms to isolate them from the background \shortcite{Medina2010}. Shadows can also result in false detections \shortcite{Rhodes2005}. 

\paragraph{Loop Detectors} 
An induction loop detector consists of a loop of wire embedded in the pavement, which is actuated when a vehicle passes over it. Loop detectors are more robust to the environment than cameras. However, actuation depends on the detector's sensitivity, which is hard to configure accurately and may result in undercounting or overcounting. Overcounting can also occur due to excessive sensitivity to adjacent lanes (splashover) or detector interference (chattering) \shortcite{Lee2012}.

\section{Demand Modeling Pipeline}
\label{sec:pipeline}

\subsection{Computer Vision-Based Vehicle Counting from Camera Footage}
\label{sec:pipeline:cv}
\begin{figure}[ht]
    \centering
    \includegraphics[trim=0cm 13.33cm 0cm 0cm,clip,width=0.8\linewidth]{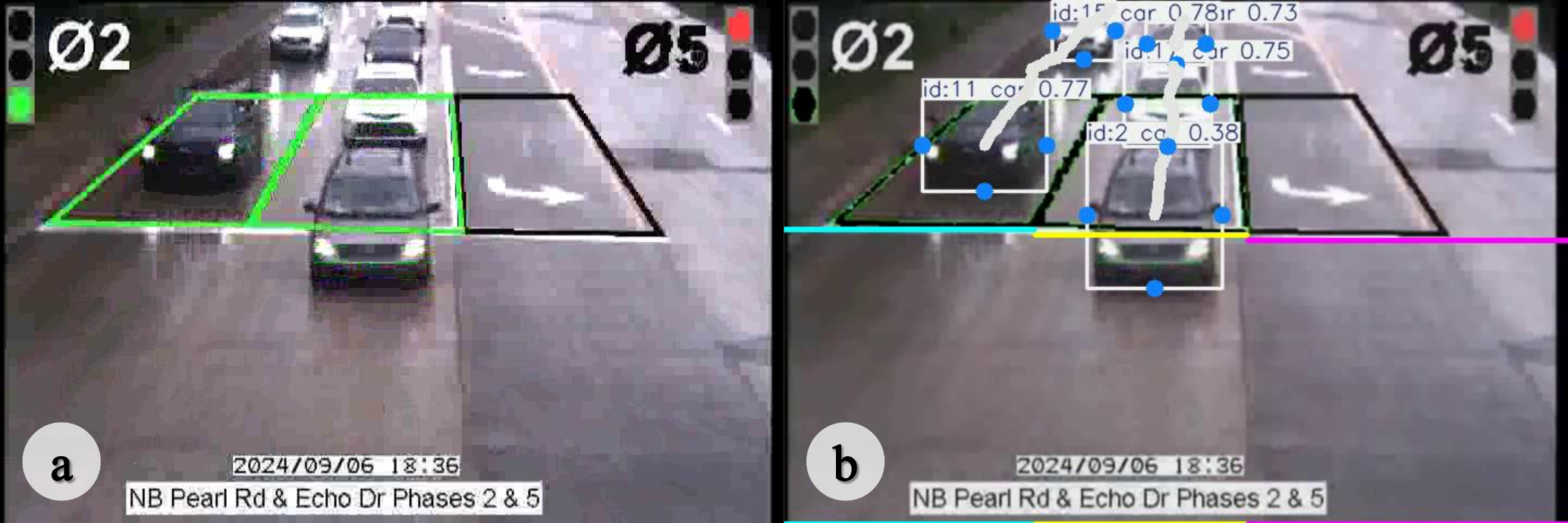}
    \caption{Demonstration of our vehicle tracking method on camera detector footage from intersection 8 (US 42 \& Echo Rd) in Strongsville, Ohio. (a) Raw footage, showing two vehicles actuating a detection zone in the center lane. (b) Footage with preprocessing filters applied, and bounding boxes and tracks for counted vehicles annotated. The colored lines represent manually labeled stop bar positions.}
    \label{fig:cv}
\end{figure}

To detect individual vehicles more accurately than the camera detectors themselves, we use the YOLOv11 object detection model \shortcite{Jocher2024}. In each frame of footage, the model predicts a set of bounding boxes, each of which encloses an object. For each box, the model outputs a class $b_c$, the $x$ and $y$ coordinates of the center of the bounding box ($b_x$ and $b_y$), and the width and height of the box ($b_w$ and $b_h$).  

How can YOLO's detections of vehicles be converted to counts for each lane? We manually annotate each frame of the traffic footage with the stop bar's $y$-coordinate ($S_y$), and with $x$-coordinates for each lane ($S_L^\ell, S_R^\ell$). The most straightforward method to perform counting is to verify whether the bounding box for an object identified as a vehicle has crossed the position of the stop bar, i.e., we increment the count for lane $\ell$ when a detected vehicle has $S_L^\ell \leq b_x \leq S_R^\ell$, and $|b_y - S_y| \leq \epsilon$ for some predefined threshold $\epsilon$. 

When we apply this method directly in practice, we encounter two issues. (1) Due to instability in the real-time streaming (RTSP) connection over which detector footage is retrieved, frames are frequently dropped. For many vehicles, this leads to the absence of the frames in which their bounding boxes' borders $b_y$ are close to the stop bar $S_y$. These vehicles first appear far above the stop bar ($b_y \ll S_y$), and then far below the stop bar ($b_y \gg S_y$) once the footage resumes. (2) Pixelation artifacts, particularly around the detection zones marked on the footage, also obfuscate the bounding boxes.

To address dropped frames, we use the BoT-SORT algorithm \shortcite{Aharon2022} to perform vehicle tracking. BoT-SORT reidentifies each bounding box across consecutive frames to provide a consistent ID $b_t$. In each frame, for each bounding box, we verify whether $b_t$ has already been counted. To ensure that we do not capture traffic in other directions, we only consider $b_t$ if $b_y < S_y$ initially. If $b_t$ has not yet been counted and $b_y > S_y$, we increment the vehicle count and mark $b_t$ as counted. Even if the footage is missing the frame where $|b_y - S_y| \leq \epsilon$ for a vehicle, the vehicle will be counted in a later frame. We also apply filters to smooth the footage as a preprocessing step, including a non-local means filter \shortcite{Buades2011}, a spatiotemporal denoising filter (\texttt{hqdn3d}), a frame-blending motion interpolation filter (\texttt{minterpolate}), and a filter to remove detector actuation overlays. The results are shown in \Cref{fig:cv}b.

\subsection{Optimization-Based Vehicle Route Generation from Multimodal Data}
\label{sec:pipeline:qp}
Our vehicle tracking-based computer vision method from \Cref{sec:pipeline:cv} outputs vehicle counts $f^{CV}_j$ for a set of counting locations $j$. These represent the traffic volumes at the eastbound, northbound, southbound, and westbound approaches for each intersection (if they are available). We also have a set of counts from loop detectors $f^{LD}_j$, which overlap with the computer vision counts at a subset of counting locations. But how can these multimodal vehicle counts be integrated to generate vehicle routes for a fully-realized simulation?

We make two key assumptions to identify the set of feasible routes. (1) Given an origin and destination in the road network, we assume that vehicles perform shortest-path (Dijkstra) routing. (2) We assume that most of the traffic in the road network originates at the fringes of the network, and few routes begin and end in the middle of a road edge (representing traffic from unmodeled driveways). This assumption holds as long as all major sources and sinks of traffic are modeled. Based on these assumptions, we enumerate the full set of routes between all counting locations, instead of randomly sampling them as in prior work.

Given this set of feasible routes, we aim to solve for the number of times each route should be used, so that the number of times they pass through the counting locations match the given vehicle counts as closely as possible. We do so by dividing 24 hours of count data into 15-minute time segments, which are indexed as $t \in \{0..95\}$. For each time segment, we match the total counts at each counting location, as well as counts for dedicated left-turn and right-turn lanes if they exist.

How closely should we match the counts? We assume that, for some counting locations $j$ and time segments $t$, we have ground truth counts $f^M_{jt}$, and that error exists in both our computer vision counts $f^{CV}_{jt}$ and loop detector counts $f^{LD}_{jt}$. Let $M$, $CV$, and $LD$ denote the sets of counting locations for these three sources. Based on how much computer vision overcounts or undercounts for locations in common with the ground truth, we extrapolate conservative lower and upper bounds for the true counts:
\begin{align*}
    \alpha^{CV}_{\mathrm{LB}} = \min_t \min_{j \in M \cap CV} \frac{f^M_{jt}}{f^{CV}_{jt}}, \alpha^{CV}_{\mathrm{UB}} = \max_t \max_{j \in M \cap CV} \frac{f^M_{jt}}{f^{CV}_{jt}}.
\end{align*}
We also derive bounds for loop detectors, $\alpha^{LD}_{\mathrm{LB}}$ and $\alpha^{LD}_{\mathrm{UB}}$, in a similar fashion. Then, we assume that, for time segments $t \in \{0..95\}$, $\alpha^{CV}_{LB} f^{CV}_{jt} \leq f^M_{jt} \leq \alpha^{CV}_{UB} f^{CV}_{jt}, \forall j \in CV$, and $\alpha^{LD}_{LB} f^{LD}_{jt} \leq f^M_{jt} \leq \alpha^{LD}_{UB} f^{LD}_{jt}, \forall j \in LD$.

Now, for each 15-minute time segment $t$, we used the solver Gurobi to solve the following quadratic integer program (QIP), where the decision variable is the number of usages $r_{it}$ for each route $i \in \{1, \ldots, n\}$:
\begin{subequations}
\label{eq:qp}
\begin{align}     
    \min_{\mathbf{r}_t,\mathbf{s}^{CV}_t,\mathbf{s}^{LD}_t}\ &\lVert{\mathbf{s}^{CV}_t}\rVert^2_2 +  \lVert{\mathbf{s}^{LD}_t}\rVert^2_2 + \lambda_\textrm{nonfringe} \sum_{i \in \mathrm{nonfringe}} r_{it} + \lambda_\textrm{temporal} \lVert{\mathbf{r}_t - \mathbf{r}_{t-1}}\rVert^2_2 \label{qp:obj} \\
    \textrm{s.t.}\ & \mathbf{A}\mathbf{r}_t + \mathbf{s}^{CV}_t \geq \alpha^{CV}_{LB} \mathbf{f}^{CV}_t \label{qp:cv-lb} \\
    & \mathbf{A}\mathbf{r}_t + \mathbf{s}^{CV}_t \leq \alpha^{CV}_{UB} \mathbf{f}^{CV}_t \label{qp:cv-ub} \\
    & \mathbf{A}\mathbf{r}_t + \mathbf{s}^{LD}_t \geq \alpha^{LD}_{LB} \mathbf{f}^{LD}_t \label{qp:ld-lb} \\
    & \mathbf{A}\mathbf{r}_t + \mathbf{s}^{LD}_t \leq \alpha^{LD}_{UB} \mathbf{f}^{LD}_t \label{qp:ld-ub} \\
    & \mathbf{r}_t \in {(\mathbb{Z}^{\geq 0})}^n \notag,  
\end{align}
\end{subequations}
where $\mathbf{A} \in \{0,1\}^{n \times m}$ is a binary matrix denoting which counting locations are used by routes: $\mathbf{A}_{ij}$ is 1 if route $i$ passes counting location $j$, and is 0 otherwise, such that $\mathbf{A}\mathbf{r}_t$ gives the number of times the generated routes collectively pass each counting location $j \in \{1, \ldots, m\}$; $\mathbf{s}^{CV}_t, \mathbf{s}^{LD}_t \in \mathbb{R}^m$ are slack variables that represent the error between the generated routes' counts $\mathbf{A}\mathbf{r}_t$ and the actual counts $\mathbf{f}^{CV}_t$ and $\mathbf{f}^{LD}_t$; nonfringe is the set of indices for routes where the start or the end of the route are interior edges in the road network; $\lambda_\textrm{nonfringe}$ is a hyperparameter for weighting the objective function penalty for these routes; and $\lambda_\textrm{temporal}$ is a hyperparameter for penalizing discrepancies between the generated routes of adjacent time segments.

In the QIP, constraints \eqref{qp:cv-lb} and \eqref{qp:cv-ub} specify that $\mathbf{A}\mathbf{r}_t$ should lie within a probable range of counts extrapolated from computer vision counts. The lower bound $\alpha^{CV}_{LB} \mathbf{f}^{CV}_t$ assumes that computer vision is overcounting, and the upper bound $\alpha^{CV}_{UB} \mathbf{f}^{CV}_t$ assumes that it is undercounting. The sum-of-squares of the error $\lVert{\mathbf{s}^{CV}_t}\rVert^2_2$ is minimized in the objective function \eqref{qp:obj}. The two following constraints, \eqref{qp:ld-lb} and \eqref{qp:ld-ub}, are analogous constraints for loop detector counts. Again, the error $\lVert{\mathbf{s}^{LD}_t}\rVert^2_2$ is minimized in the objective. Because all of these bounds may be relatively loose, the problem is underconstrained. Within the possible space of solutions, the final two terms in the objective function optimize for two heuristics of simulation realism: first, the $\lambda_\textrm{nonfringe}$ term minimizes the number of nonfringe routes, which are rare under our assumptions; and second, the $\lambda_\textrm{temporal}$ term encodes the intuition that traffic flow should be relatively continuous over time.

The solution $\mathbf{r}$ represents a set of vehicles that should enter the road network within the 15-minute time segment indexed as $t$. How can these vehicles be distributed within the time segment? The simplest strategy is to uniformly distribute them. However, under this strategy, two adjacent traffic segments with similar volumes may have very different traffic patterns. Instead, we solve the following QP that enforces similarity between the traffic patterns of adjacent time segments:
\begin{align*}
    \min_{\mathbf{c}_t}  & \lVert{\mathbf{c}_t - \mathbf{c}_{t-1}}\rVert^2_2 \\
    \mathrm{s.t.}\ & \sum_{m=1}^{15} c_{i,(t,m)} = r_{it}, \forall i \in \{1, \ldots, n\}, \\
    & \mathbf{c}_t \in (\mathbb{Z}^{\geq 0})^{n \times 15}
\end{align*}
where $\mathbf{c}_t$ is the minute-by-minute distribution of routes $\mathbf{r}_t$ within time segment $t$: $c_{i,(t,m)}$ is the number of times a vehicle with route $i$ appears in the simulation during minute $m$ of the 15-minute time segment $t$. As a base condition, we uniformly distribute traffic for the first time segment $t = 0$, but we solve this QP for time segments $t \in \{1, \ldots, 95\}$. 

More sophisticated methods could be used to enforce temporal continuity in traffic flow. Instead of a uniform distribution, the knowledge of stakeholders could be used to initialize the traffic flow of $t = 0$. Dynamics models could also be used to compute the expected travel time of vehicles through the road network. This would mitigate cases where a vehicle appears in time segment $t$, but does not reach an intersection where it is counted in $\mathbf{f}_t$ until time segment $t + 1$ or later.

\subsection{LLM Agent Simulation Refinement from Natural Language Feedback}
\label{sec:pipeline:llm}
QIP \eqref{eq:qp} is fundamentally underconstrained.  For each counting location, the generated counts could lie anywhere between the lower bounds (constraints \eqref{qp:cv-lb} and \eqref{qp:ld-lb}) and the upper bounds (constraints \eqref{qp:cv-ub} and \eqref{qp:ld-ub}). Additionally, most municipalities do not install camera detectors at every intersection, meaning that our vehicle tracking method (\Cref{sec:pipeline:cv}) does not generate counts or impose bounds for the entire road network. Not all possible ways of assigning routes to match these counts are equally realistic. We can leverage the domain knowledge of stakeholders, such as traffic engineers, to ensure that the traffic simulation is aligned with downstream use cases such as traffic analysis. Yet, without experience in optimization, it is difficult for these stakeholders to directly modify QIP \eqref{eq:qp} to align with their intuition.

Our problem formulation is as follows. We are given $K$ pieces of structured natural language feedback $p^{(k)} = (t^{(k)}, j^{(k)}, l^{(k)})$, where each piece consists of a time, intersection, and a natural language description of what corrections (if any) should be made to the simulated traffic state at this intersection. We are also given code which solves the original problem $\mathrm{QP}^{(0)}$, and the route counts $\mathbf{A}\mathbf{r}^{(0)}$ obtained from solving $\mathrm{QP}^{(0)}$. The objective is to produce an updated problem $\mathrm{QP}^{(K)}$, which has been modified so that it will produce a new route set $\mathbf{r}^{(K)}$ that addresses $\{p^{(1)}, \ldots, p^{(K)}\}$. The core difficulty in this problem is converting the natural language feedback into concrete optimization constraints, which cannot be accomplished by traditional optimization methods. Instead, we solve this problem by using an LLM agent to answer prompts containing $(p^{(k)}, \mathrm{QP}^{(k-1)}, \mathbf{A}\mathbf{r}^{(k-1)})$, and leveraging its code generation capabilities to generate $\mathrm{QP}^{(k)}$.

Notably, we do not provide the LLM agent with any handcrafted information beyond the time segment $t^{(k)}$ and intersection $j^{(k)}$ that the feedback is targeted at; what is already available from $\mathrm{QP}^{(k)}$; and a list of intersections and main roads. Based on the set of route counts $\mathbf{A}\mathbf{r}^{(k-1)}$ from the previous simulation, the LLM agent must automatically extract concrete, quantitative constraints that are aligned with the qualitative feedback. To solve this task, we prompt the LLM agent using a chain of thought \shortcite{Wei2022} to:
\begin{enumerate}[(1)]
    \item Extract the relevant counts by formulating a call to a \texttt{get\_counts} tool, which retrieves the previous counts $\mathbf{A}\mathbf{r}^{(k-1)}_{jt}$ for a particular counting location and time segment $(j, t)$;
    \item Write a constraint corresponding to the feedback $p^{(k)}$, using the counts from the previous step to make subjective judgments on how to set undetermined coefficients;
    \item Translate this constraint to Python code for the package \texttt{cvxpy} \shortcite{Diamond2016};
    \item For the time segment $t^{(k)}$ specified in the feedback, add this constraint to the optimization function while minimally modifying the rest of the code;
    \item For adjacent time segments $(t^{(k)} - 1, t^{(k)} + 1)$, add relaxed constraints to ensure temporal continuity.
\end{enumerate}

We use this LLM agent within an automated, iterative simulation refinement framework like that of \shortciteN{Behari2024}. For each of $K$ pieces of feedback, we first use the LLM agent to generate a program. Then, we apply a rapid verification procedure to the generated program based on three criteria: 
\begin{enumerate}[(1)]
    \item \emph{Syntactic correctness}. We attempt to execute the program in a Python interpreter to ensure it represents syntactically correct Python. If not, then it cannot generate an updated simulation.
    
    \item \emph{Feasibility}. We attempt to solve the new QIP for the time segment $t^{(k)}$, as well as for adjacent time segments $(t^{(k)} - 1, t^{(k)} + 1)$. Assuming that the feedback is internally consistent, and given the underconstrained nature of the problem, we expect that the solver should be able to quickly find (see runtime results in \Cref{sec:results:qp}) at least one feasible solution $\hat{\mathbf{r}}^{(k)}$ for the LLM-generated QIP.
    
    \item \emph{Semantic correctness}. We attempt to verify that the LLM agent's modification to the simulation actually corresponds to the feedback given, based on the solution $\hat{\mathbf{r}}^{(k)}$ to the feasibility check. To do so, we use the LLM agent to perform \emph{self-reflection} \shortcite{Shinn2023}. It uses the \texttt{get\_counts} tool to first retrieve relevant counts $\mathbf{A}\mathbf{r}^{(k-1)}$ from the previous solution, and then the same counts $\mathbf{A}\hat{\mathbf{r}}^{(k)}$ from the candidate solution. Then, the LLM compares these counts while taking into account the feedback $p^{(k)}$ to return a binary signal of whether the modification is semantically correct.
\end{enumerate}

If the program fails any of these three verification criteria, we discard the program and prompt the LLM agent to generate a new one. This process repeats until the LLM agent generates a correct program $\mathrm{QP}^{(k)}$ for feedback $p^{(k)}$. In the next iteration, we prompt the LLM agent to directly modify $\mathrm{QP}^{(k)}$ to produce $\mathrm{QP}^{(k+1)}$. After at least $K$ generations, we obtain a single program $\mathrm{QP}^{(K)}$, which we execute for all time steps $t$ to obtain the final solution $\mathbf{r}^{(K)}$ and a corresponding simulation.

\section{Simulation Results: Strongsville, Ohio}
\label{sec:results}
We applied our demand modeling pipeline from \Cref{sec:pipeline} to simulate a large road network in the city of Strongsville, Ohio. The Strongsville road network experiences heavy through traffic due to its connection to two interstates, I-71 and I-80; the ramps of these interstates respectively connect to two intersecting arterials, SR~82 (Royalton Road) and US~42 (Pearl Road). The daily traffic volumes of both of these roads have exceeded their designed capacities, leading to the implementation of various countermeasures to improve throughput \shortcite{Euthenics2023}. As part of these countermeasures, Strongsville installed an adaptive traffic signal control system on SR~82 and US~42. 

This system uses three types of detectors to adjust the signaling pattern for its controllers. (1) Camera detectors are used for the main roads at each intersection (i.e., along SR~82 and US~42) and on some side roads. (2) Loop detectors are used for most side roads and some turning movements. (3) Radar detectors are used for detection upstream and downstream of intersections. As our goal is to match the traffic state at the intersections themselves, we do not consider data from Strongsville's radar detectors.

\begin{figure}[ht]
    \centering
    \includegraphics[width=0.625\linewidth]{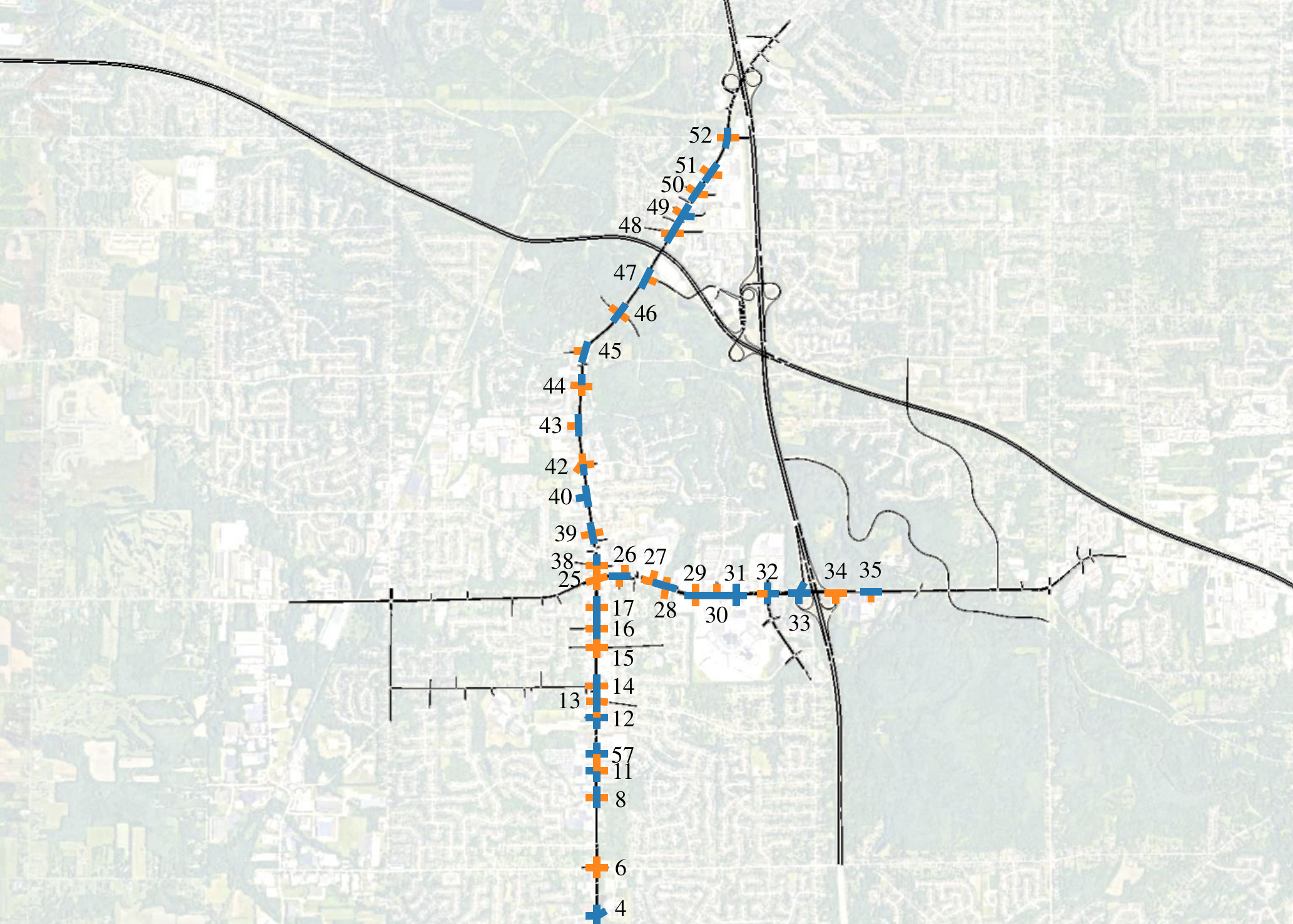}
    \caption{Screenshot of traffic simulation of Strongsville, Ohio. Counts are available for 36 intersections, either derived from vehicle tracking on raw camera detector footage (approaches in blue) or from loop/AutoScope detectors (approaches in orange). Each intersection is labeled with its number.}
    \label{fig:simulation}
\end{figure}

Our simulation, as shown in \Cref{fig:simulation}, covers the intersections along SR~82 and US~42 for which the city has installed adaptive signal control. We first converted OpenStreetMap data to a SUMO \shortcite{AlvarezLopez2018} road network. Next, on Friday, September 6, 2024, we captured 24 hours of footage from 74 out of 86 counting locations where AutoScope detectors are installed. We used counts from AutoScope and loop detectors to fill in missing counts from vehicle tracking. After we applied our pipeline, we randomly assigned vehicles to different vehicle classes \shortcite{Weinblatt2013}, following a survey conducted by the Ohio DOT in September 2022. Finally, we implemented the traffic signal patterns that were in use.

In the rest of this section, we evaluate the accuracy of our pipeline steps for this simulation.

\subsection{Accuracy of Vehicle Counting}
\label{sec:results:cv}
To evaluate the accuracy of our vehicle tracking-based counting method (\Cref{sec:pipeline:cv}), we manually counted traffic from camera detector footage. As doing so would be infeasible for the entire simulation, we selected footage from four different intersections that are important to stakeholders (from the south, center, east, and north of the road network), and two one-hour time segments (12 pm, an off-peak hour, and 5 pm, a peak hour) for each intersection. \Cref{fig:cv} shows a screenshot from one of these pieces of footage.

\begin{figure}[ht]
    \centering
    \includegraphics[width=0.8\linewidth]{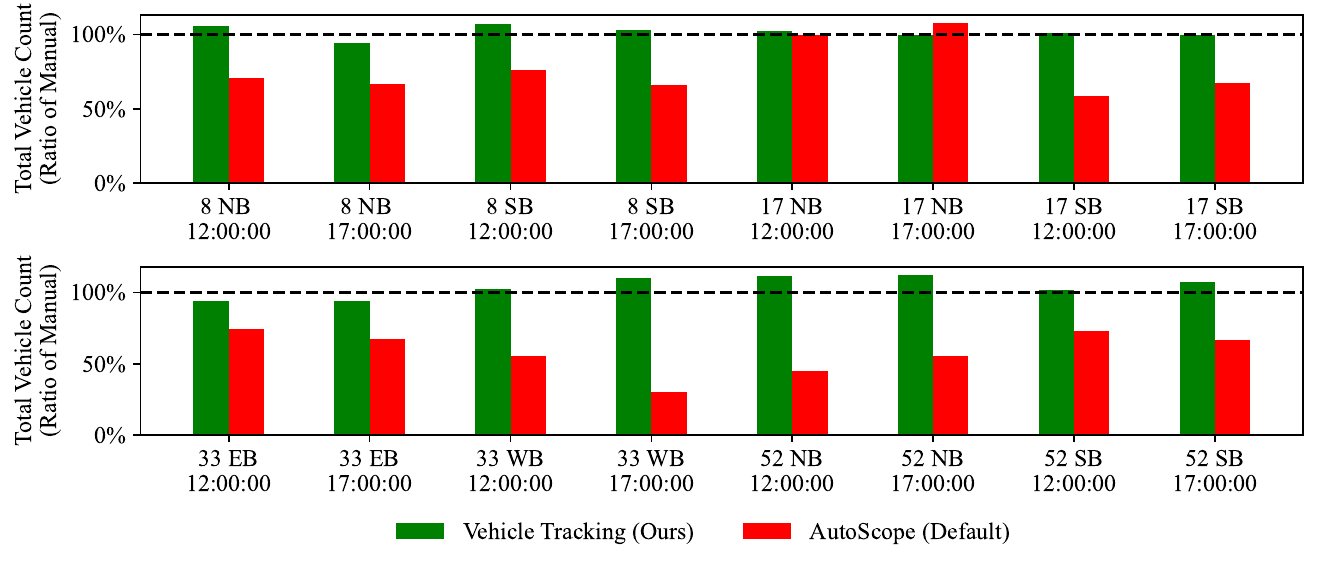}
    \caption{Plot of counts from our vehicle tracking method (green) and AutoScope (red), as ratios relative to manual counts. If a method's result perfectly matches the manual counts, it has a ratio of 100\%.}
    \label{fig:cv-eval}
\end{figure}

\begin{table}[htb]
\centering
\caption{Comparison of vehicle counts from our method (CV) and AutoScope (AS) with manual counting.}
\begin{tabular}{lcccccccc}
\toprule
& \multicolumn{4}{c}{8 --- Pearl Rd \& Echo Rd} & \multicolumn{4}{c}{17 --- Pearl Rd \& Business Entrance} \\
\cmidrule(lr){2-5} \cmidrule(lr){6-9}
& \multicolumn{2}{c}{Northbound} & \multicolumn{2}{c}{Southbound} & \multicolumn{2}{c}{Northbound} & \multicolumn{2}{c}{Southbound} \\
\cmidrule(lr){2-3} \cmidrule(lr){4-5} \cmidrule(lr){6-7} \cmidrule(lr){8-9}
& 12:00 & 17:00 & 12:00 & 17:00 & 12:00 & 17:00 & 12:00 & 17:00 \\
\midrule
Manual  & 823 & 959 & 781 & 1041 & 1042 & 952 & 1106 & 1329 \\
\midrule
CV    & 872 & 905 & 839 & 1071 & 1065 & 950 & 1114 & 1322 \\
AS      & 582 & 637 & 594 &  688 & 1036 & 1026 & 647 & 895 \\
\midrule
& \multicolumn{4}{c}{33 --- Royalton Rd \& I-71 Ramps} & \multicolumn{4}{c}{52 --- Pearl Rd \& Sprague Rd} \\
\cmidrule(lr){2-5} \cmidrule(lr){6-9}
& \multicolumn{2}{c}{Eastbound} & \multicolumn{2}{c}{Westbound} & \multicolumn{2}{c}{Northbound} & \multicolumn{2}{c}{Southbound} \\
\cmidrule(lr){2-3} \cmidrule(lr){4-5} \cmidrule(lr){6-7} \cmidrule(lr){8-9}
& 12:00 & 17:00 & 12:00 & 17:00 & 12:00 & 17:00 & 12:00 & 17:00 \\
\midrule
Manual  & 1889 & 2088 & 1059 &  726 & 979 & 1048 & 1166 & 1441 \\
\midrule
CV    & 1779 & 1962 & 1089 & 801 & 1096 & 1178 & 1191 & 1553 \\
AS      & 1405 & 1406 &  587 &  220 & 442 & 578 & 853 & 960 \\
\bottomrule
\end{tabular}
\label{tab:cv-eval}
\end{table}

In \Cref{fig:cv-eval}, we compare the counts generated by our vehicle tracking method and by AutoScope to the ground truth from manual counting. Based on our evaluation, our method was able to faithfully capture the traffic state of Strongsville. For all 16 of the footage excerpts that we manually counted, our method had an error of less than 2 vehicles per minute (120 vehicles per hour). We used the same set of hyperparameters for preprocessing and detection/tracking across all counting locations; tuning these hyperparameters for individual counting locations could yield further gains. 

Meanwhile, AutoScope exhibited a persistent pattern of undercounting across all of the approaches that we manually counted. In fact, it had an error of \emph{more} than 3 vehicles per minute (180 vehicles per hour) for 14 out of 16 footage excerpts, with the exception being intersection 17's northbound approach (where we observed that lane switching resulted in duplicated actuations). The primary cause of this undercounting was the continuous actuation of detection zones by consecutive vehicles, as we discussed in \Cref{sec:detectors}. Consistent with this, AutoScope was generally more accurate under intermittent traffic during the 12 pm time segment, and its accuracy degraded under increased traffic levels during the 5 pm time segment.

We used these results to estimate lower and upper bound ratios between our vehicle tracking counts and the ground truth: $\alpha^{CV}_{\mathrm{LB}} = 0.94, \alpha^{CV}_{\mathrm{UB}} = 1.12$. As there is insufficient overlap between manual and loop detector counts, we extrapolated bounds for loop detector counts based on the ratio between vehicle tracking and loop detector counts. These bounds are loose due to the level of error in the loop detectors:
\smallskip
\begin{align*}
    \alpha^{LD}_{\mathrm{LB}} = \alpha^{CV}_{\mathrm{LB}} \min_t \min_{j \in CV \cap LD} \frac{f^{CV}_{jt}}{f^{LD}_{jt}} = 0.02, \alpha^{LD}_{\mathrm{UB}} = \alpha^{CV}_{\mathrm{UB}} \max_t \max_{j \in CV \cap LD} \frac{f^{CV}_{jt}}{f^{LD}_{jt}} = 19.06.
\end{align*}

\subsection{Accuracy of Generated Simulation}
\label{sec:results:qp}
Next, we solved QIP \eqref{eq:qp} to generate a set of routes consistent with these counts (\Cref{sec:pipeline:qp}). Owing to the underconstrained nature of the problem, obtaining a feasible solution for the 18\,496-variable QIP in each time segment required less than 0.5 seconds; thus, our approach scales well to moderately-sized road networks. However, to optimize solution quality in the final simulation, we ran the QIP for each time segment for 60 seconds. The objective function's value was mainly determined by bound violations for vehicle tracking counts $\lVert{\mathbf{s}^{CV}_t}\rVert^2_2$, which were two orders of magnitude larger than violations for loop detector counts $\lVert{\mathbf{s}^{LD}_t}\rVert^2_2$ and the fringe route penalty $\sum_{i \in \mathrm{nonfringe}} r_{it}$. To balance the objective function, we set the hyperparameter $\lambda = 10$. We obtained a simulation with a total volume of 182\,230 vehicles over 24 hours. Among these vehicles, 72.64\% had routes that started and ended on the fringes of the road network.

\begin{figure}[ht]
    \centering
    \includegraphics[width=0.8\linewidth]{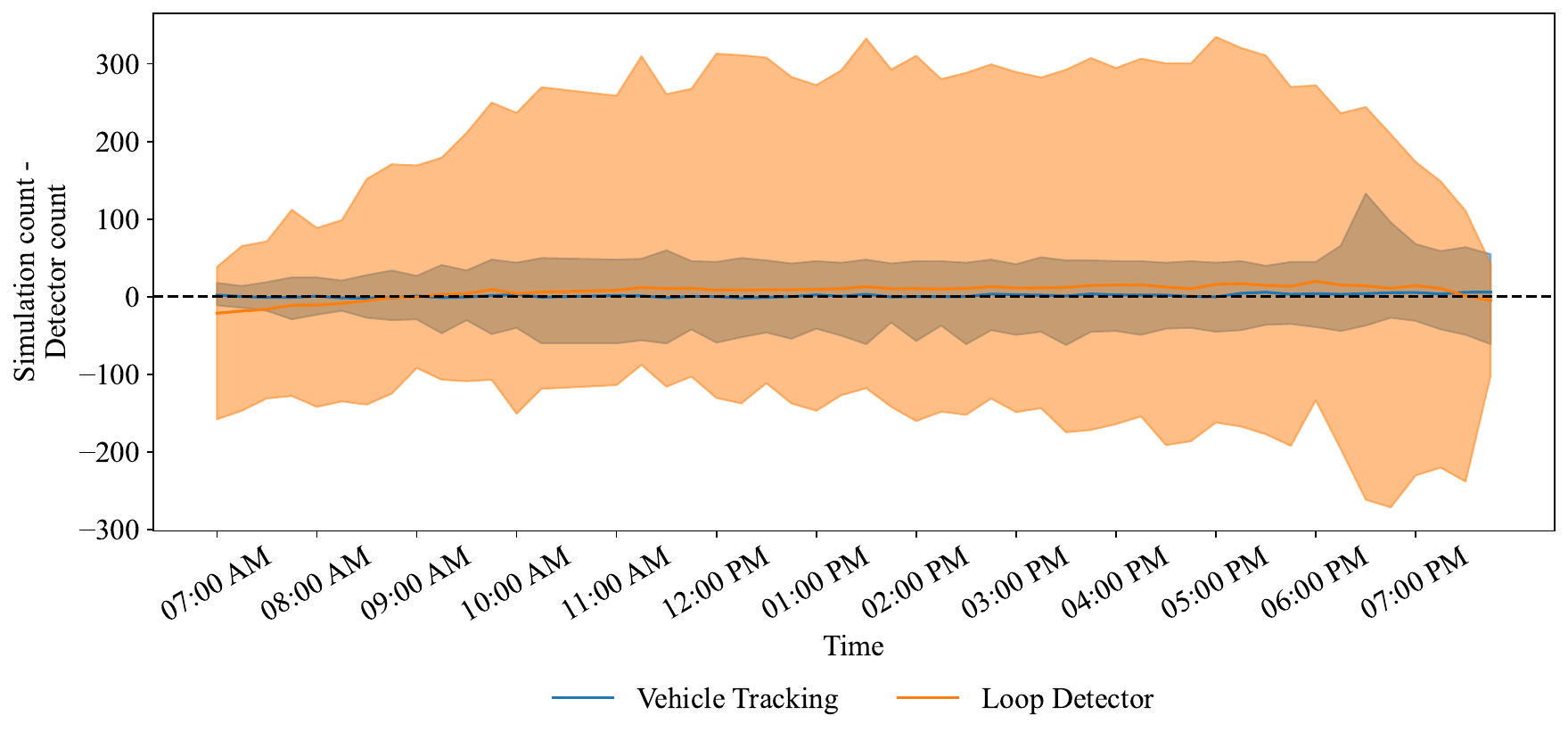}
    \caption{Plot of difference between counts in the QIP-generated simulation, counts from our vehicle tracking-based computer vision method, and counts from loop detectors. For each time segment, the solid line represents the mean across counting locations, while the shaded region represents the range.}
    \label{fig:range-eval}
\end{figure}

In \Cref{fig:range-eval}, we focus on the accuracy of our simulation for the time interval between sunrise (6:59 am) and sunset (7:51 pm) on September 6, 2024. Across counting locations on average, the simulation was accurate to our computer vision counts $f^{CV}_j$, with no overflow or underflow. This can be attributed to the relatively narrow range of $[\alpha^{CV}_{\mathrm{LB}}, \alpha^{CV}_{\mathrm{UB}}]$. However, there are individual counting locations where the simulation has substantial overflow or underflow, especially so for the loop detector counts. We attribute these to counting locations with few vehicles where the detected traffic flow is inconsistent. Violating the expected bounds of these counts results in a small penalty compared to the rest of the objective function.

\subsection{Accuracy of LLM Agent-Generated Constraints}
\label{sec:results:llm}
As our LLM for iterative simulation refinement (\Cref{sec:pipeline:llm}), we used Claude 3.7 Sonnet. An earlier iteration of this model achieved state-of-the-art performance on code generation benchmarks \shortcite{Zhuo2025}. We performed two rounds of evaluation: one on synthetically-generated feedback (where a ground truth exists for constraint correctness), and one on real stakeholder feedback (where there is no ground truth). Our evaluation focused on the three criteria used by the LLM agent to verify generated code (\Cref{sec:pipeline:llm}): syntactic correctness, feasibility, and semantic correctness.

First, we randomly generated $K = 20$ pieces of structured feedback in the form of (intersection, direction, approach, increase/decrease) tuples. We used Claude to rephrase this feedback to match the style of stakeholder feedback. Here, we did not use reflection to assess semantic correctness, but instead directly verified the traffic volume in the updated simulation against the structured feedback. We generated ten programs for each piece of feedback with a temperature of 0.8. 

\begin{figure}[ht]
    \centering
    \includegraphics[width=\linewidth]{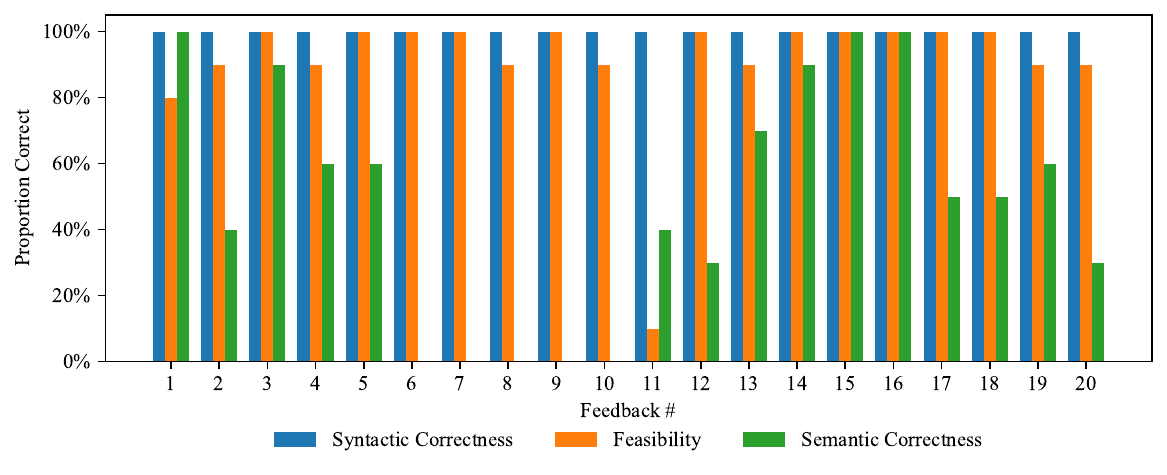}
    \caption{Evaluation of $B = 10$ optimization programs generated by Claude for $K = 20$ synthetically generated pieces of feedback. We used three criteria to evaluate the programs: syntactic correctness (perfect), feasibility, and semantic correctness (both high but not perfect).}
    \label{fig:prompt-eval}
\end{figure}

As shown in \Cref{fig:prompt-eval}, the LLM agent always generated valid Python code, giving a syntactic correctness rate of \textbf{100\%}. The feasibility rate was \textbf{87\%}. We found that tool use was important to prevent hallucination of counts. Not all generated programs were feasible due to two issues in the added constraints: (1) they included a slack variable term $s_t$, which conflicted with the slack variable constraints from the original optimization problem, or (2) they were formulated in terms of vehicle tracking-based computer vision counts $f_t^{CV}$, which were not always available. Lastly, the semantic correctness rate was \textbf{87\%} --- whenever the generated program was feasible, the result was also correct. This gave us confidence in continuing to use our approach, after adding the reflection procedure and modifying the prompt to prevent the two aforementioned issues.

Second, we collected $K = 20$ pieces of feedback by presenting our simulation for the 5:00 pm time segment to a stakeholder familiar with Strongsville's traffic conditions. Among the 20 pieces of feedback, 12 pointed to intersections that were true to real life (particularly those with counts based on vehicle tracking-based computer vision), while 8 pointed to intersections where the simulated traffic needed improvement. With reflection in place, we generated a single program for each piece of feedback, again with a temperature of 0.8. The LLM agent had a syntactic correctness rate of \textbf{100\%}, a feasibility rate of \textbf{100\%}, and a semantic correctness rate of \textbf{100\%} from reflection, including for the 8 pieces of feedback that indicated changes. 

Now, we demonstrate the capabilities of our framework with examples of the constraints generated by the LLM agent. Intersection 25 (US 42 \& SR 82) was an intersection with camera detectors for which we were unable to capture footage. The stakeholder commented:
\begin{quote}
    I would expect to see more cars both eastbound and westbound at this time of day. Especially at this intersection at this time of day, I would expect every approach to max out. Each left turn, each through, main line, side street, I would expect it to be pretty packed at this time of day. There might be just a couple of lingering [vehicles], but I'd expect pretty much each queue to get its full allocated time.
\end{quote}

The LLM agent queried the total eastbound, northbound, southbound, and westbound counts for the old simulation during the 5:00 pm time segment (119, 157, 281, and 334), as well as turn counts for the eastbound and westbound approaches. It then generated the following code:
\begin{verbatim}
# Add constraints for intersection 25 during peak hour (5:00 PM)
# Indices for intersection 25: 40 (EB), 41 (NB), 42 (SB), 43 (WB)
if hour == "05:00 PM":
    # Enforce higher traffic at intersection 25 based on 
    # traffic engineer's feedback
    # Increase eastbound traffic (was 119)
    flow_constraints.append((cat_locs[:, 40, 0].T @ route_count) 
        >= 200)
    # Increase westbound traffic (was 334)
    flow_constraints.append((cat_locs[:, 43, 0].T @ route_count) 
        >= 450)
    # Increase northbound traffic (was 157)
    flow_constraints.append((cat_locs[:, 41, 0].T @ route_count) 
        >= 250)
    # Increase southbound traffic (was 281)
    flow_constraints.append((cat_locs[:, 42, 0].T @ route_count) 
        >= 400)

    # Also increase left and right turn volumes
    # Eastbound left turns (was 7)
    flow_constraints.append((cat_locs[:, 40, 1].T @ route_count) 
        >= 30) 
    # Eastbound right turns (was 43) 
    flow_constraints.append((cat_locs[:, 40, 2].T @ route_count) 
        >= 80) 
    # Westbound left turns (was 178)
    flow_constraints.append((cat_locs[:, 43, 1].T @ route_count) 
        >= 250)
    # Westbound right turns (was 138)
    flow_constraints.append((cat_locs[:, 43, 2].T @ route_count) 
        >= 200)
    # Northbound right turns (was 108)
    flow_constraints.append((cat_locs[:, 41, 2].T @ route_count) 
        >= 150)
        
    # Southbound left turns (was 141)
    flow_constraints.append((cat_locs[:, 42, 1].T @ route_count)
        >= 200)
    # Southbound right turns (was 17) 
    flow_constraints.append((cat_locs[:, 42, 2].T @ route_count) 
        >= 40) 
    
# Add interpolated constraints for adjacent time segments 
# (4:45 PM and 5:15 PM) 
elif hour == "04:45 PM" or hour == "05:15 PM":
    # Apply milder increases (about 75% of the 5:00 PM increase) 
    # for adjacent time periods
    # Eastbound traffic
    flow_constraints.append((cat_locs[:, 40, 0].T @ route_count) 
        >= 180)
    # Westbound traffic
    flow_constraints.append((cat_locs[:, 43, 0].T @ route_count) 
        >= 400)  
    # Northbound traffic
    flow_constraints.append((cat_locs[:, 41, 0].T @ route_count) 
        >= 220)
    # Southbound traffic
    flow_constraints.append((cat_locs[:, 42, 0].T @ route_count) 
        >= 350)
\end{verbatim}
where \texttt{cat\_locs} is the counting location matrix $\mathbf{A}$, and \texttt{route\_count} is the solution $\mathbf{r}_t^{(k)}$.

This snippet demonstrates that the LLM agent is able to correctly identify the indices of the counting locations (40 to 43); quantify the stakeholder's expectation of additional traffic by adding lower bound constraints on the total flow and on the turning flows; and add looser constraints to the adjacent time segments for continuity, all in valid Python code. 

For other pieces of feedback where the stakeholder indicated that the simulation was accurate, the LLM agent added constraints to reinforce the existing traffic patterns in the simulation. For intersection 46 (Pearl Rd \& Valley View Pkwy), the stakeholder commented:
\begin{quote}
    There's not a whole lot of traffic there on the side streets, what you have going there looks accurate.
\end{quote}
This led to the following code, which imposes both lower and upper bounds on traffic volumes:
\begin{verbatim}
# Add constraints for intersection 46 (Valley View Pkwy) 
# based on traffic engineer's feedback
# Intersection 46 corresponds to index 28, 
# so counting locations are 112-115
if hour == "05:00 PM":
    # Maintain low side street volumes 
    # as engineer confirmed they're accurate
    # Eastbound constraint (location 112)
    flow_constraints.append(cat_locs[:, 112, 0].T @ route_count 
        >= 45)  # Lower bound
    flow_constraints.append(cat_locs[:, 112, 0].T @ route_count 
        <= 57)  # Upper bound

    # Westbound constraint (location 115)
    flow_constraints.append(cat_locs[:, 115, 0].T @ route_count 
        >= 42)  # Lower bound
    flow_constraints.append(cat_locs[:, 115, 0].T @ route_count 
        <= 52)  # Upper bound

# Add similar but relaxed constraints for adjacent time periods
elif hour == "04:45 PM" or hour == "05:15 PM":
    # Slightly more relaxed bounds for adjacent time periods
    # Eastbound constraint (location 112)
    flow_constraints.append(cat_locs[:, 112, 0].T @ route_count 
        >= 40)  # Lower bound
    flow_constraints.append(cat_locs[:, 112, 0].T @ route_count 
        <= 62)  # Upper bound
    # Westbound constraint (location 115)
    flow_constraints.append(cat_locs[:, 115, 0].T @ route_count 
        >= 37)  # Lower bound
    flow_constraints.append(cat_locs[:, 115, 0].T @ route_count 
        <= 57)  # Upper bound
\end{verbatim}

Our final simulation of Strongsville was created by executing the simulation refinement procedure in sequence for all $K = 20$ pieces of feedback from the stakeholder. For each modification to the original simulation, the LLM agent's reflection procedure indicated that it accurately captured the feedback. When the stakeholder was presented with the final simulation, they concurred with the LLM agent regarding the improvements that had been made. The simulation had a total volume of 200\,332 vehicles over 24 hours, with 64.72\% having fringe routes. This increase can be in part attributed to increased volume at intersections without vehicle tracking counts.

\section{Conclusion}
\label{sec:conclusion}
In this work, we presented an end-to-end pipeline for modeling demand in traffic simulations with three steps: computer vision-based vehicle counting, combinatorial optimization-based vehicle route generation, and LLM-based iterative simulation refinement from natural language feedback. We applied our pipeline to a high-traffic road network in Strongsville, Ohio. Based on our evaluation results, our demand modeling methodology adheres more faithfully to real-world traffic conditions than approaches used in past work, and it holds promise in generalizing to road networks from other municipalities with similarly multimodal detector data. For Strongsville, we could generate simulations quickly even when we exhaustively enumerated the route set. However, the number of routes increases exponentially with the number of intersections. To improve the scalability of our pipeline for even larger road networks, we suggest that the route set should not be sampled, but instead clustered into geographic subregions connected at boundaries by major roads.

Our pipeline represents an initial, offline proof-of-concept of how LLM agents can enable interactive simulation generation. One line of future work is to convert it into a streaming pipeline capable of near-real-time use. Streaming capabilities would allow simulations to be updated based on live traffic. They would also enable the creation of an interactive interface where stakeholders can iterate on detection and optimization parameters using natural language feedback, while reviewing the results of their feedback instantaneously. We also envision that other sources of data could be incorporated into our demand modeling framework. Road state reports (e.g. Waze), weather data, and business information can all be indicative of factors that impact traffic. As LLMs' capabilities improve, they hold promise for integrating data from these heterogeneous sources \shortcite{Chang2024}. This frontier of AI-enabled possibilities can help traffic simulations to better reflect heterogeneity in real-world traffic conditions and to better serve their users.

\section*{Acknowledgments}
We thank Scott Morse and Dave Palmer of Path Master, Kyle Love and Eric Raamot of Econolite, Sean Fitzgerel of PTV Group, Michael Schweikart of TMS Engineers, Ken Mikula and Lori Daley of the City of Strongsville, and Lisa Kay Schweyer, Stan Caldwell, and Karen Lightman of Traffic21/Safety21 --- crucial stakeholders without whom this work would not have been possible. Additionally, we thank Jenny T. Liang, Kush Jain, Sean Qian, Naveen Raman, Yixuan Xu, and Jingwu Tang for their invaluable ideas and feedback on the technical portions of this work. This work was supported by a research grant from Mobility21, a US DOT National University Transportation Center, and the Tang Family Endowed Innovation Fund. 

\footnotesize

\bibliographystyle{wsc}

\bibliography{ref}

\section*{Author Biographies}
\noindent \textbf{REX CHEN} (email \email{rexc@cmu.edu}) is a PhD graduate from the Societal Computing program of the Software and Societal Systems Department in the School of Computer Science at Carnegie Mellon University. His research focuses on applying reinforcement learning and other AI techniques to transportation and other socially impactful domains. His work aims to design AI systems capable of addressing the key deployment considerations of stakeholders in these domains.
\newline

\noindent \textbf{KAREN WU} (email \email{karenw2@andrew.cmu.edu}) is a second-year undergraduate student at the School of Computer Science at Carnegie Mellon University.
\newline

\noindent \textbf{JOHN MCCARTNEY} (email \email{john.mccartney@pathmasterinc.com}) is a 2002 University of Toledo graduate with a B.S. in Mechanical Engineering Technology. John is a Systems Support Engineer with 23 years in the transportation industry and 20 years at Path Master, which is a distributor of traffic control equipment for Ohio, Western Pennsylvania, Kentucky, and West Virginia. He is responsible for providing technical support and training for traffic controllers, detection, and management systems to industry professionals. His day-to-day activities include overseeing the implementation of Centracs ATMS and Mobility systems in the Path Master territory, troubleshooting software, electrical, and communication issues, plus guiding traffic signal projects to completion.
\newline 

\noindent \textbf{FEI FANG} (email \email{feifang@cmu.edu}) is an Associate Professor in the Software and Societal Systems Department in the School of Computer Science at Carnegie Mellon University. Her research interests lie in the area of artificial intelligence and multi-agent systems, focusing on the integration of computational game theory and machine learning to address real-world challenges in critical domains such as security, sustainability, and mobility.
\newline

\noindent \textbf{NORMAN SADEH} (email \email{sadeh@cs.cmu.edu}) is a Professor in the School of Computer Science at CMU, where he is affiliated with the Software and Societal Systems Department, the Human-Computer Interaction Institute, and the CyLab Security and Privacy Institute. His research interests span mobile computing, the IoT, cybersecurity, privacy, machine learning, AI, and related public policy issues. His past work includes deployed planning and scheduling technologies for commercial systems.

\end{document}